\documentclass[11pt]{article}

\usepackage[utf8]{inputenc}
\usepackage[english]{babel}
\usepackage{color}

\usepackage{xr}

\usepackage{CJKutf8}
\usepackage{titling}

\usepackage{amsfonts,amsmath,amssymb,amsbsy}
\usepackage{mathpazo}
\usepackage{mathptmx}

\usepackage{graphicx}
\usepackage{pdflscape}

\usepackage{tabularx}
    \newcolumntype{L}{>{\raggedright\arraybackslash}X}
\usepackage{dcolumn} 
\usepackage{booktabs}
\usepackage{threeparttable}
\usepackage{comment}
\usepackage{multirow}

\usepackage[justification=justified,singlelinecheck=false]{caption}
\usepackage{subcaption}
\usepackage{float}
\restylefloat{table}
\restylefloat{figure}

\usepackage{xcolor}
\usepackage{footnotebackref}
\usepackage{hyperref}
\hypersetup{
    colorlinks=true,
    citecolor=black,
    linkcolor=black,
    urlcolor=black
}
\usepackage{enumitem}

\usepackage[square,numbers]{natbib}

\makeatletter
\newcommand*{\addFileDependency}[1]{
  \typeout{(#1)}
  \@addtofilelist{#1}
  \IfFileExists{#1}{}{\typeout{No file #1.}}
}
\makeatother

\newcommand*{\myexternaldocument}[1]{%
    \externaldocument{#1}%
    \addFileDependency{#1.tex}%
    \addFileDependency{#1.aux}%
}

\myexternaldocument{som2} 

\usepackage{setspace}

\usepackage{geometry}
\usepackage{afterpage}

\evensidemargin\oddsidemargin
\title{Semantic Processing of Political Words in Naturalistic Information Differs by Political Orientation}\author{Shuhei Kitamura\thanks{%
Center for Infectious Disease Education and Research, Osaka University, D74-1 Office for Industry-University Co-Creation (Bldg. D), 2-8 Yamadaoka, Suita, Osaka, 565-0871 Japan. To whom correspondence may be addressed. Email: kitamura@cider.osaka-u.ac.jp.} \and Aya S. Ihara\thanks{Center for Information and Neural Networks, Advanced ICT Research Institute, National Institute of Information and Communications Technology, Kobe, 651-2492, Japan}}
\date{}

\begin{document}

\maketitle

\begin{abstract}
Worldviews may differ significantly according to political orientation. Even a single word can have a completely different meaning depending on political orientation. However, no direct evidence has been obtained on differences in the semantic processing of single words in naturalistic information between individuals with different political orientations. The present study aimed to fill this gap. We measured electroencephalographic signals while participants with different political orientations listened to naturalistic content. Responses for moral-, ideology-, and policy-related words between and within the participant groups were then compared. Within-group comparisons showed that right-leaning participants reacted more to moral-related words than to policy-related words, while left-leaning participants reacted more to policy-related words than to moral-related words. In addition, between-group comparisons also showed that neural responses for moral-related words were greater in right-leaning participants than in left-leaning participants and those for policy-related words were lesser in right-leaning participants than in neutral participants. There was a significant correlation between the predicted and self-reported political orientations. In summary, the study found that people with different political orientations differ in semantic processing at the level of a single word. These findings have implications for understanding the mechanisms of political polarization and for making policy messages more effective. \\
\noindent {\bf Classification:} Biological Sciences; Psychological and Cognitive Sciences \\
\noindent {\bf Keywords:} Electroencephalography, political orientation, conservatives, liberals.
\end{abstract}

\clearpage

\noindent {\large {\bf Significance Statement.}} People's political attitudes are becoming more polarized globally, especially in the United States. Do people with different political orientations understand the same language differently, even at the level of a single word? In this study, we measured neural activity while participants listened to naturalistic content. The results showed that participants with different political orientations comprehended political words differently, even at the level of a single word. This result suggests that the information processing bias that drives polarization could occur at the level of single words. \\

\noindent {\bf Author Contributions.} S.K. and A.S.I. designed the study and prepared the experiment. S.K. collected data. A.S.I. performed the EEG analysis. S.K. and A.S.I. performed the statistical analysis. S.K. and A.S.I. wrote the paper. \\

\noindent {\bf Competing Interest Statement.} The authors declare no competing interests. \\

\section{Introduction}

In the recent years, people’s political views have become more polarized globally, especially in the United States \citep{Pew_2014, Boxell_2022}. Polarization is also evident in party manifestos \citep{Gethin_2021}, with voters’ political views becoming more aligned with those of the party they support \citep{Gentzkow_2016}. Although this is a serious social problem, the mechanisms underlying these phenomena are not fully understood \citep{baar_2022}.

In the United States, there are known partisan differences in campaign speeches and political advertising \citep{westen_2007, lakoff_2008, haidt_2012}. Republican slogans, political commercials, and speeches “go straight for the gut,” while those of the Democrats tend to emphasize specific policies and the benefits they will bring to voters \citep[][p.181]{haidt_2012}. An interpretation of such a difference is that Republicans appeal more to the voters’ cognitive {\it intuition (automatic processes)}, while Democrats appeal more to their cognitive {\it reasoning (controlled processes)} \citep{haidt_2012}. This is consistent with the literature on cognitive reflection that shows that liberals are more analytical than conservatives, while conservatives are more intuitive than liberals \citep{talhelm_2014, deppe_2015, yilmaz_2016, jost_2017}. Partisan differences in political language were also confirmed in studies using the phrases used by members of Congress \citep{gentzkow_2010, jensen_2012, gentzkow_2019}.

While previous studies have highlighted important partisan differences in the {\it production} of language, it remains unclear whether partisan differences in the {\it comprehension} of language also exist. As the standard Bayesian model of information processing suggests, different prior beliefs would produce different posterior beliefs from the exact same information \citep{Gerber_1999}. Previous studies of motivated reasoning (or biased assimilation) \citep{kunda_1990} have shown that people with different prior beliefs tend to interpret scientific evidence differently \citep{lord_1979, kunda_1987, plous_1991, miller_1993, kuhn_1996, munro_1997, taber_2006, corner_2012}. Consistent with these findings, Leong et al. (2020) used functional magnetic resonance imaging (fMRI) to demonstrate that brain activity diverged between conservative and liberal participants while watching the same videos related to immigration policy \cite{leong_2020}. This neural polarization, as termed by the authors, was found in the dorsomedial prefrontal cortex, and increased with the use of risk-related and moral-emotional language in the videos. The results may indicate that semantic processing differs by political orientation. In a related fMRI study, de Bruin et al. (2023) found that the intersubject similarity in political ideology is predicted by the intersubject similarity in the behavioral spatial arrangement of political words and the neural representation patterns measured while reading them \citep{bruin_2023}. In particular, similar neural activity was found for words related to the political categories ``immigrant" within the striatum and ``America" within the temporoparietal junction. In addition, using fragments of a video clip on immigration at approximately 10-s intervals, the study found significant associations between temporal neural state changes along party lines and the emotional and semantic disagreement between Democrats and Republicans that was rated by independent coders based on the content of the video clip.

These results indicate that differences in semantic processing at the level of a single word between individuals with different political orientations may exist. In addition, differences in semantic processing may also depend on differences in word categories. Therefore, the present study exploits the high temporal resolution of electroencephalography (EEG) and focuses on a component of an event-related potential (ERP) that is specifically related to semantic processing to assess directly whether real-time semantic processing at the level of a single word in naturalistic information differs by the political orientation. Furthermore, this study also examines whether such differences arise differently for certain word categories.

\begin{figure}[H]
    \begin{center}
    \includegraphics[width=0.8\textwidth]{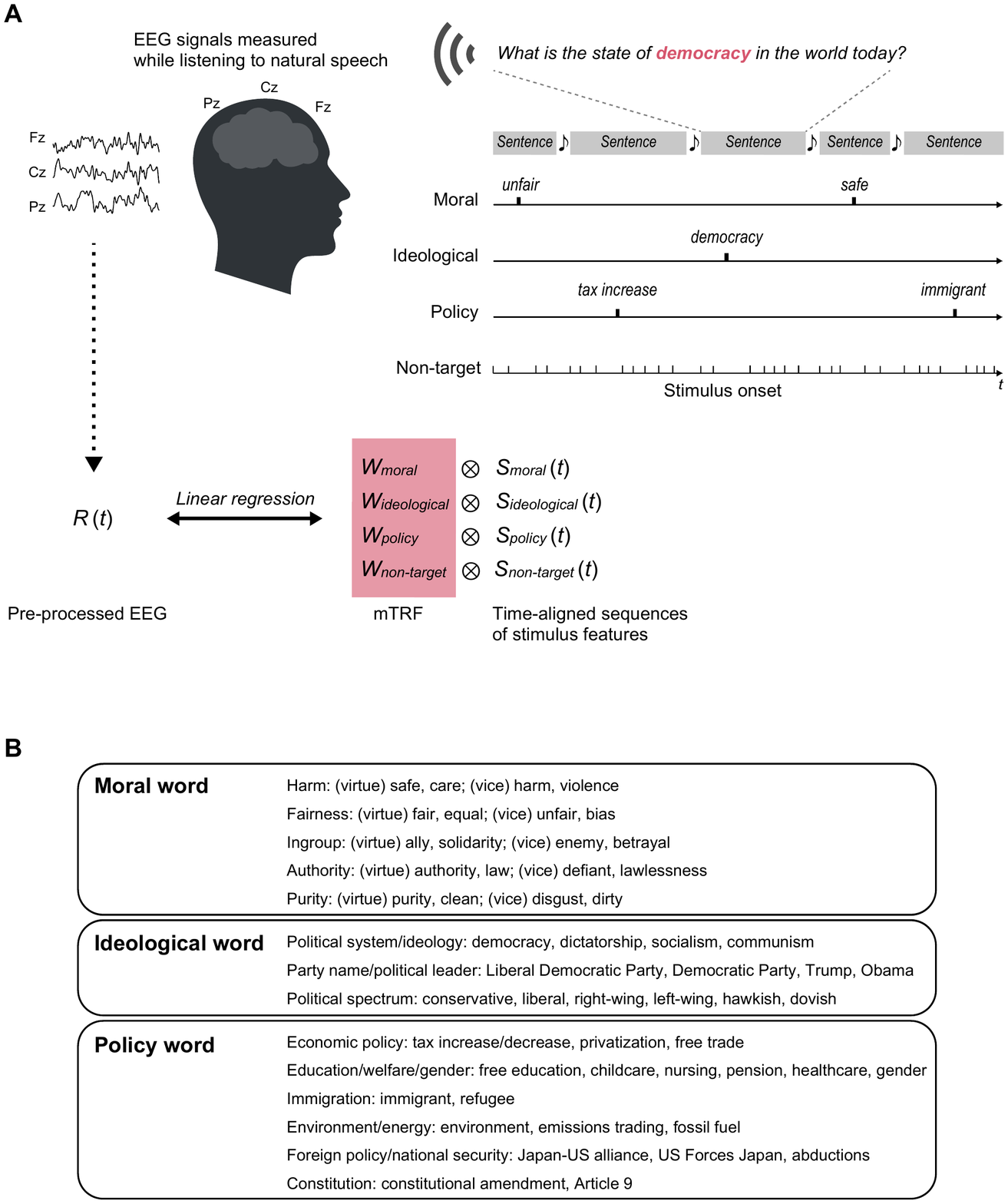}
    \vspace{-2cm}
    \caption{Summary of the experiment and analysis. (A) EEG signals were measured while participants with different political orientations listened to naturalistic content, including a target word that belonged to one of the following three categories of political words: moral, ideological, and policy. A multivariate temporal response function (mTRF) that describes the linear mapping between ongoing stimuli (word onsets) and ongoing EEG signals was estimated. (B) Samples of moral, ideological, and policy words were shown.\label{fig:studydesign}}
    \end{center}
\end{figure}

EEG signals were recorded in 35 young adults who were divided into three groups based on their self-reported political orientation: political left (liberal), neutral, and right (conservative) groups. For this study, N400, a component of an ERP, could suit as an index because it is associated with semantic processing. Its amplitude is modulated by the ease of access to long-term memory and integration with the preceding context \citep[for reviews,][]{kutas_2000, kutas_2011}. However, with conventional methods, ERPs for words in natural speech are generally difficult to extract due to an overlapped response with a small interval between words. Therefore, we estimated the temporal response function (TRF) that describes the linear mapping between ongoing stimuli (word onsets) and ongoing EEG signals \citep{crosse_2016} (Figure \ref{fig:studydesign}A). Previous studies showed a component corresponding to N400 in the TRF weight time locked to word onset within natural speech \citep{broderick_2018, ihara_2021, fuseda_2022}. 

Using this method, the present study compared N400 as a neural response to the following three types of political words (Figure \ref{fig:studydesign}B) in natural speech across three groups of individuals with different political orientations. First, {\it moral words}, such as harm and fair, based on the Moral Foundations Dictionary were included \citep{graham_2009}. Using words from the same dictionary, Graham et al. (2009) found that liberal and conservative preachers use different moral words in church sermons \cite{graham_2009}. Second, {\it ideological words}, related to political system and ideology (e.g., dictatorship, communism), the name of a political party and a political leader (e.g., Democratic Party, Trump), and political spectrum (e.g., conservative) were included. These words were meant to capture political ideology in various dimensions. Finally, {\it policy words} surrounding policies (e.g., free trade) were included. We considered that these three types of political words were ordered from more fundamental political words (moral words) to more specific political words (policy words). To use real-world political content, the original text of the audio stimuli was taken from a non-partisan news commentary television program and recorded by a professional narrator. If conservatives produced political language that appealed to people’s cognitive intuition (automatic processes) and liberals produced those that appealed to cognitive reasoning (controlled processes) \citep{haidt_2012}, we might also observe similar partisan differences in the comprehension of language. Assuming that moral words are more related to automatic processes while policy words are more related to controlled processes, we hypothesized that the politically right-leaning group would respond more to moral words than to policy words and the left-leaning group would respond more to policy words than to moral words. Similarly, we also hypothesized that neural responses to moral words would be greater in the right-leaning group than those in the left-leaning group, while those to policy words would be lesser in the right-leaning group than those in the left-leaning group. Finally, we also predicted the individual’s political orientation using the EEG responses.

\section{Results}

\subsection{Correlations between Political Orientation and Attitudes toward Policies.}
To evaluate the validity of our grouping of participants based on their self-reported political orientation, we compared it with individual attitudes toward various policies. A total of 19 major policies were selected using questionnaires from previous studies which, as the present study does, asked participants about their political preferences \citep{oxley_2008, ahn_2014}, and using the most common political survey questionnaire in Japan, that is, the UTokyo-Asahi Survey. The UTokyo-Asahi Survey was used to select policies particularly relevant in the Japanese context. Of these 19 policies, 12 were taken from prior literature and the rest were taken from the second source. Participants evaluated each policy using a 4-point scale, with 1 indicating disagreement and 4 indicating agreement. The average values were then compared between the left- and right groups ({\it SI Appendix}, Figure \ref{fig:poli_attitudes}). The Wilcoxon rank sum test was performed for each outcome. The results showed significant between-group differences in attitudes toward 11 of the 19 policies. Specifically, compared with the left group, the right group was more likely to disagree with accepting immigrants ($p$ = 0.007), allowing foreigners to vote ($p$  = 4.015 $\times$ 10$^{-4}$), a female emperor ($p$ = 0.018), removing US bases in Japan ($p$ = 0.004), separate surnames ($p$ = 0.021), social security provided by the government ($p$ = 0.001), and socialism ($p$ = 0.012). Meanwhile, they were more likely to agree with capital punishment ($p$ = 0.007), politicians visiting Yasukuni Shrine ($p$ = 0.002), revising Article 9 of the Constitution ($p$ = 0.004), and strengthening the military ($p$ = 0.001). Finally, there were no significant differences in attitudes toward ethical education ($p$ = 0.396), gender equality ($p$ = 0.425), government regulations ($p$ = 0.369), nuclear energy/power plants ($p$ = 0.164), same-sex marriage ($p$ = 0.350), small government ($p$ = 0.116), tax reduction ($p$ = 0.382), and trade liberalization ($p$ = 0.906).

\begin{figure}
    \begin{center}
    \includegraphics[width=0.9\textwidth]{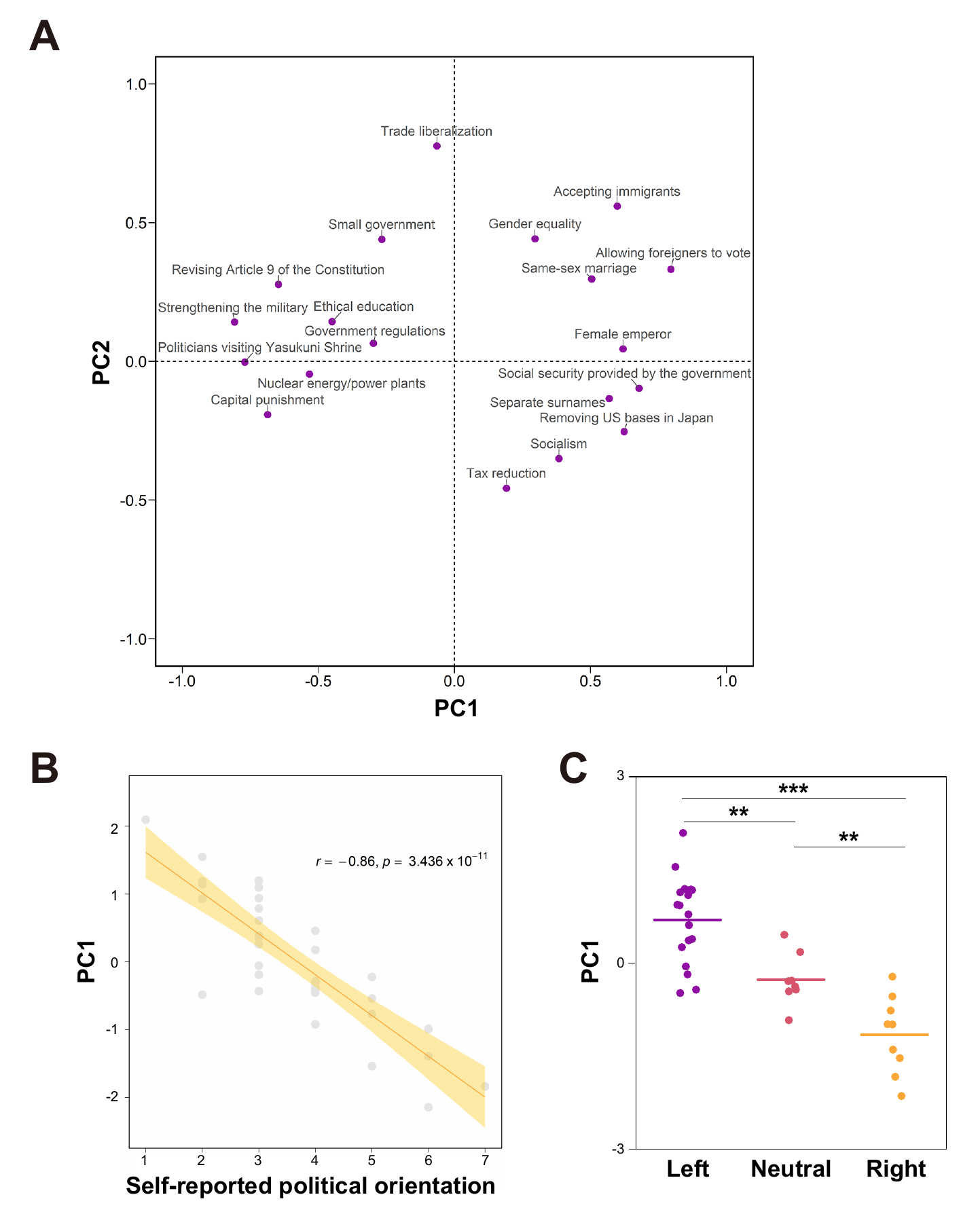}
    \caption{Validation of our grouping of participants based on their self-reported political orientations. (A) A principal component analysis (PCA) was performed on the participants’ evaluation of the 19 policies, and maps of the first and second principal components (PC1 and PC2, respectively) were presented. (B) The PC1 scores showed a significant correlation with the self-reported political orientation, which ranged from 1 (political left or liberal) to 7 (political right or conservative). (C) The PC1 scores were significantly different between the left and neutral groups, between the left and right groups, and between the neutral and right groups. **$p$ $<$ 0.01 and ***$<$ 0.001.\label{fig:group}}
    \end{center}
\end{figure}

Next, a principal component analysis (PCA) for participants' evaluations of the 19 policies was performed. The map with the first and second principal components (PC1 and PC2, respectively) are shown in Figure \ref{fig:group}A. Pearson correlation analysis revealed that PC1 was significantly correlated with self-reported political orientation ($r$ = -0.86, $p$ = 3.436 $\times$ 10$^{-11}$) (Figure \ref{fig:group}B). In addition, the results of the one-way analysis of variance (ANOVA) showed that the PC1 scores differed significantly among the three groups ($F$(2,34) = 27.39, $p$ = 1.167 $\times$ 10$^{-7}$): left $>$ neutral ($p$ = 0.003), left $>$ right ($p$ = 8.121 $\times$ 10$^{-8}$), and neutral $>$ right ($p$ = 0.004) (Figure \ref{fig:group}C). These results indicate that our grouping of participants, based on their self-reported political orientation, was valid.

\subsection{People with Different Political Orientations Reacted Differently to Each Category of Words.}
\begin{figure}
\begin{center}
\includegraphics[width=0.9\linewidth]{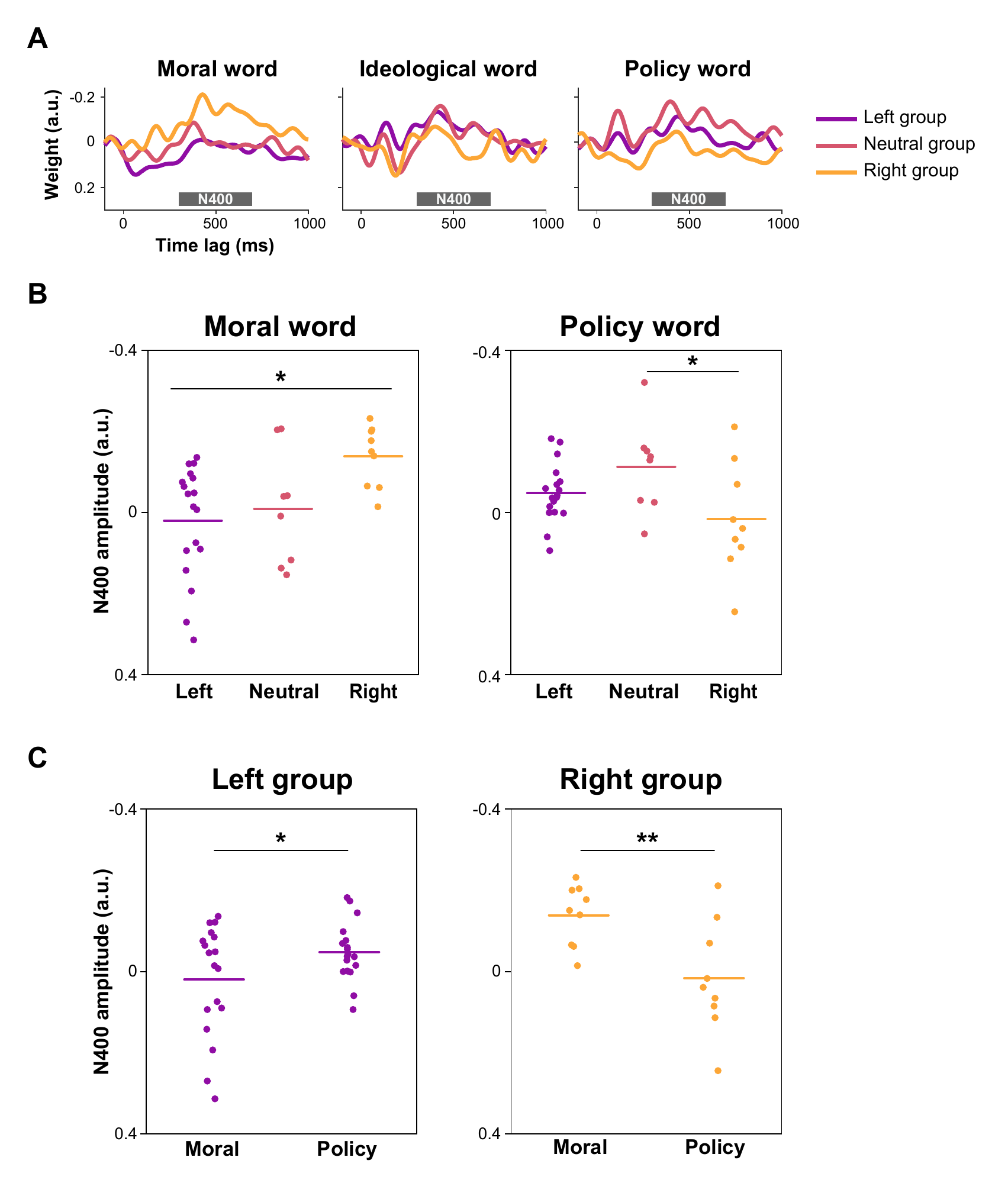}
\caption{Differences in N400 between and within groups. (A) Grand-averaged TRF weights for each word category are shown. (B) On one hand, the N400 amplitude for moral words was negatively larger in the right group than in the left group. On the other hand, the N400 amplitude for policy words was negatively smaller in the right group than in the neutral group. (C) The left group showed a greater N400 amplitude for policy words than for moral words, while the right group elicited a greater N400 amplitude for moral words than for policy words. *$p$ $<$ 0.05 and **$<$ 0.01.\label{fig:N400}}
\end{center}
\end{figure}

Grand-averaged TRF weights for each word category are shown in Figure \ref{fig:N400}A. To clarify the effect of political orientation on semantic processing for political words when listening to naturalistic content, we performed a two-way mixed-design ANOVA, using a within-subject factor of Word (moral, ideological, and policy) and a between-subject factor of Group (left, neutral, and right), for the amplitude of N400. A significant interaction of Word $\times$ Group was found ($F$(4, 64) = 4.192, $p$ = 0.004, $\eta_p^2$ = 0.208), and no significant main effects of Word ($F$(2, 64) = 0.022, $p$ = 0.978, $\eta_p^2$ = 0.001) and Group ($F$(2, 62) = 0.384, $p$ = 0.684, $\eta_p^2$ = 0.023) were found. A significant main effect of Group was found for moral words ($F$(2, 34) = 4.482, $p$ = 0.015) and policy words ($F$(2, 34) = 3.458, $p$ = 0.044), while there was no significant main effect for ideological words ($F$(2, 34) = 0.970, $p$ = 0.390). The N400 amplitude for moral words was negatively larger in the right group than in the left group ($p$ = 0.013, Figure \ref{fig:N400}B). In contrast, the N400 amplitude for policy words was negatively smaller in the right group than in the neutral group ($p$ = 0.039). Furthermore, the planned comparison revealed that the right group elicited a greater N400 amplitude for moral words than for policy words (t(8) = ‒3.605, $p$ = 0.004, Figure \ref{fig:N400}C); in contrast, the left group had a greater N400 amplitude for policy words than for moral words (t(17) = 2.248, $p$ = 0.038). 

The summary of the results is as follows. The within-group comparison revealed that the right group responded more to moral words than to policy words, while the left group responded more to policy words than to moral words. In addition, the between-group comparison revealed that neural responses to moral words were greater in the right group than in the left group and those to policy words were smaller in the right group than in the neutral group. These results were consistent with our hypotheses, except that we did not find a significant statistical difference between the right and left groups, but instead found a significant statistical difference between the right and neutral groups, in terms of policy words.

\subsection{Neural Responses Predict Political Orientations.} 
\begin{figure}
    \begin{center}
    \includegraphics[width=0.8\textwidth]{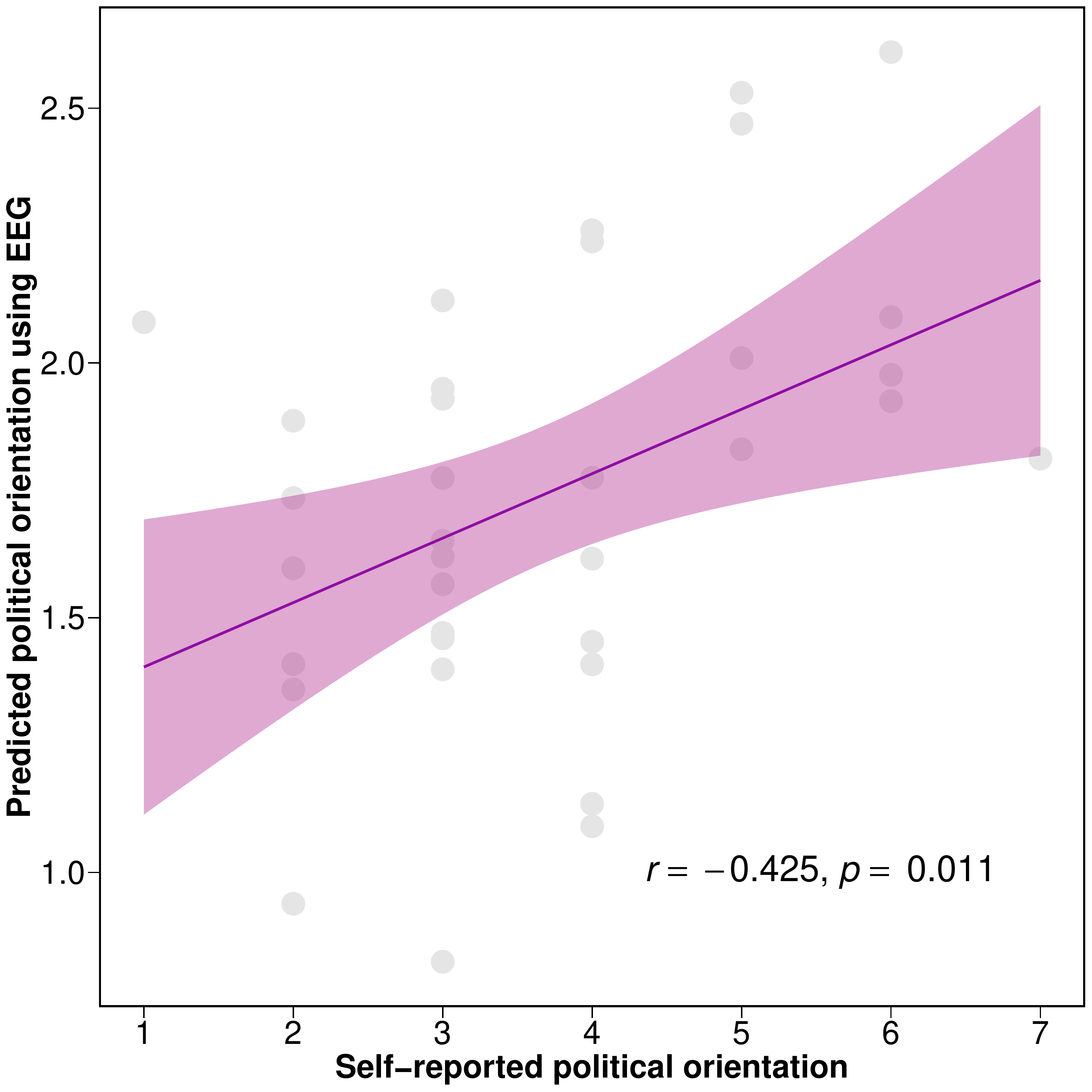}
    \caption{Predicted versus self-reported political orientations. The political orientation for each individual was predicted after regressing equation (\ref{eq:regression}) using ordinary least squares. The predicted values were plotted against the self-reported values.\label{fig:prediction_orig}}
    \end{center}
\end{figure}

Next, we used ordinary least squares to predict individual political orientations based on the neural responses to political words, and examined the correlation between the predicted values and the self-reported political orientations. There was a significant correlation between the predicted values and the self-reported values ($r$ = 0.425, $p$ = 0.011, Figure \ref{fig:prediction_orig}).

\section{Discussion}

This study found that semantic processing of the same word differs by political orientation. Our hypothesis was that the politically right group would respond more to moral words than to policy words and that the politically left group would respond more to policy words than to moral words. Confirming our hypothesis, within-group comparisons showed that the right group reacted more to moral words than to policy words, while the left group reacted more to policy words than to moral words. In addition, between-group comparisons also showed that neural responses to moral words were greater in the right group than in the left group and those to policy words were smaller in the right group than in the neutral group. These are broadly consistent with previous findings that brain activity diverges between conservative- and liberal-leaning individuals while watching the same videos, and that neural polarization increases with the use of risk-related and moral-emotional language in videos \citep{leong_2020}. In addition, these results are also broadly consistent with previous findings that the intersubject similarity in representation patterns in the word reading task predicts ideological similarity for particular political categories, and that temporal state changes along party lines are associated with emotional and semantic differences between Democrats and Republicans \citep{bruin_2023}. Our results further prove that semantic processing at the level of a single word in naturalistic information differs among individuals with different political orientations. Furthermore, the present study also showed that different types of political words have different effects.

The findings could indicate that moral words may be related to cognitive automatic processes, such as intuition, and that policy words may be related to cognitive controlled processes, such as reasoning \citep{haidt_2012}. This is also consistent with findings on cognitive reflection that liberals are more analytical than are conservatives and that conservatives are more intuitive than are liberals \citep{talhelm_2014, deppe_2015, yilmaz_2016, jost_2017}. Importantly, such partisan differences may be caused by differences in brain structure and function. Previous literature found that brain structure and function differ between liberals and conservatives \citep{amodio_2007, kanai_2011}. For example, Kanai et al. (2011) found that an increased gray matter volume of the anterior cingulate cortex is associated with greater liberalism, while increased gray matter in the right amygdala is associated with greater conservatism \cite{kanai_2011}.

The present results also have an important implication for policy messaging. Scientific messaging, especially on politically sensitive topics such as vaccination and climate change, is not always effective in persuading individuals if they have strong prior beliefs. The National Academies of Sciences, Engineering, and Medicine write that one approach to avoid a direct challenge to strongly held beliefs is tailoring scientific messages \citep{nasem_2017}. Previous studies found that moral (re)framing had a strong effect on conservatives \citep{feinberg_2012, wolsko_2016}, whereas scientific framing backfired with them \citep{palm_2021}. The heterogeneity of neural responses across different political orientations found in this study indicates that policy messages may resonate more with conservatives if the messages are framed to speak to people’s intuition using more fundamental words such as moral words. Meanwhile, the messages may resonate more with liberals if the messages are framed to speak to people’s reasoning using more specific words, such as policy words and scientific facts.

Using EEG, this study found that semantic processing at the level of a single word differs among individuals with different political orientations. Future research can examine whether, based on these differences in semantic processing, posterior beliefs also diverge and, in turn, influence the polarization of political beliefs. In addition, although the present study focuses on semantics, future research can examine its relation to other information processing stages such as attention, emotion, and memory to further understand the mechanisms underlying political polarization \citep{baar_2022}.

\section{Materials and Methods\label{sec:methods}}

\subsection{Participants.} 
Thirty-eight native Japanese speakers participated in the experiment. The selection procedure is shown in the Supplementary Material ({\it SI Appendix}, Section \ref{sec:selection}). All participants were students at Osaka University at the time of the experiment. They had normal vision and hearing and no history of neurological or psychiatric disease. The data of three participants were excluded from the analysis because their EEG data were not recorded correctly. Of the remaining 35 participants (23 males, 12 females; mean age, 23.7 years $\pm$ [standard deviation (SD), 6.8 years]), 18 participants self-reported themselves as politically left leaning (11 males, 6 females, 1 queer; mean age, 25.2 years $\pm$ [SD, 9.1 years]), 8 participants as neutral (4 males, 4 females; mean age, 22.3 years $\pm$ [SD, 1.6 years]), and 9 participants as right leaning (8 males, 1 female; mean age, 21.9 years $\pm$ [SD 1.9 years]). There were no significant differences in age and gender among the three groups ({\it SI Appendix}, Table \ref{tab:balance}).

Political orientations were self-reported in the exam room using a 7-point Likert scale, with 1 indicating political left and 7 indicating political right. In the main analysis, scale points of 1–3 were defined as the {\it left group} (liberal); 4 as the {\it neutral group}; and 5–7 as the {\it right group} (conservative). To validate this measure, we also assessed significant differences in the participants’ political attitudes toward various policies (e.g., death penalty, immigration, and constitutional amendments (4-point scale from agree to disagree)) and found that the right and left groups had significant differences in their attitudes toward 11 of the 19 policies ({\it SI Appendix}, Figure \ref{fig:poli_attitudes}). Furthermore, a PCA of the attitudes to these 19 policies was performed. Then, a one-way ANOVA was applied to the scores of the major principal components (eigenvalues $\geq$ 1) to assess whether political attitudes significantly differed between the three groups. Since the main effect of Group was significant, multiple comparisons were conducted, in which {\it p}-values were adjusted using the Holm procedure.

This study was approved by the Ethics Committee of the Osaka School of International Public Policy, Osaka University and the Ethics Committee of the Institute of Social and Economic Research, Osaka University. Informed consent was obtained from all participants. 

\subsection{Materials.} 
Speech samples from a non-partisan TV show, Jiron Koron, broadcasted by a public TV company, Japan Broadcasting Corporation (NHK) between November 15, 2016, and September 28, 2018, were selected. The show was a 10-minute non-partisan program broadcasted around midnight during weekdays. News commentators, including NHK staff and university scholars, provided easy-to-understand explanations of recent news. The scripts of each program were obtained using a TV digital recorder GaraponTV\footnote{\href{https://garapon.tv/}{https://garapon.tv/}.}. The original text was split into sentences and then parsed into words with parts of speech using a common parser of Japanese text, MeCab\footnote{\href{https://taku910.github.io/mecab/}{https://taku910.github.io/mecab/}.}. 

From all the words in the data, we first selected nouns in three categories of political words (Figure \ref{fig:studydesign}B). The first word category contained moral-related words, referred to as {\it moral words}. The Japanese translated version of the Moral Foundation Dictionary was used as the basis for word selection \citep{graham_2009, matsuo_2019}. The Dictionary included five foundations (harm, fairness, ingroup, authority, and purity), and each of these contained vice and virtue words. For example, harm/virtue words included care and safe, while harm/vice words included harm and violence. Given that the purpose of this study was to elicit differential effects for the three categories of political words, the words in the five foundations were analyzed as moral words as a whole instead of further dividing them into each foundation. The second word category comprised ideology-related words, referred to as {\it ideological words}. We first set three word subcategories categorized as (a) political system and ideology (e.g., dictatorship, communism), (b) name of political party and political leader (e.g., Democratic Party, Trump), or (c) political spectrum (e.g., conservative). These subcategories aimed to capture political ideology in various dimensions. The third word category consisted of policy-related words, referred to as {\it policy words}. First, six major groups of policies were selected using the UTokyo-Asahi Survey\footnote{\href{http://www.masaki.j.u-tokyo.ac.jp/utas/utasindex\_en.html}{http://www.masaki.j.u-tokyo.ac.jp/utas/utasindex\_en.html}.}: economic policy, education/welfare/gender, immigration, environment/energy, foreign policy/national security, and constitution. Then, representative words in each group were selected from the original data. 

The words in each category were carefully selected so that the words represented each category but were also distinctive across categories. The names of political parties and political leaders in the ideological word category, and policy-related words in the policy word category, were selected so that they were also relevant in the Japanese context. In total, there were 124 unique words (moral: 50, ideological: 34, policy: 38). Owing to the limited availability of ideological and policy words, some of the target words appeared in more than one sentence, and therefore the final number of sentences was 150. The sentences were recorded by a professional female narrator.

Analysis of the familiarity of each word showed that the participants were less familiar with the ideological words than they were with the other two word types; however, they were equally familiar with the moral and policy words ({\it SI Appendix}, Figure \ref{fig:familiarity}).

\subsection{EEG experiment.}
In the EEG experiment, 150 sentences were presented in six blocks and each block consisted of five or six trials. The stimulus sequence of each trial was as follows. First, an instruction “Press” was visually presented on a liquid crystal display (EIZO Corporation, Japan) placed in front of the participant. The speech stimulus began with the participant pressing the key. The stimuli were presented binaurally through earphones (RHA Technologies Ltd., United Kingdom) (Figure \ref{fig:studydesign}A). The participants were instructed to focus on a fixation point presented in the center of the display while listening to the speech stimuli. After being presented with three to seven sentences, one of those sentences with one blank word was shown on the screen. The participants were instructed to select the appropriate word to fill in the blank from three choices offered below the sentence. They answered by pressing one of the keys. This task was set up so that participants would pay attention to the speech stimuli. The next speech sample was presented after the participant had answered or when 15 seconds had passed after the question was asked. The speech and visual stimuli were presented using Presentation, a stimulus delivery and experiment control software (Neurobehavioral Systems, Inc., USA). 

EEG and electrooculogram (EOG) signals were continuously measured throughout all blocks using an 8-channel wireless EEG device and measurement software (Polymate Mini AP108 and Mobile Acquisition Monitor 2.02, Miyuki Giken Co. Ltd., Japan). Active electrodes were placed on Fz, Cz, and Pz locations according to the International 10–20 system for EEG measurement and on the side of, and above, the left outer canthus for EOG measurement. All signals were sampled at 500 Hz using the left earlobe as the ground and the right earlobe as the online reference. Audio stimuli were prepared stereophonically; one channel contained the speech stimuli to be presented to the participants. The other contained a square wave as a trigger to indicate the onset of each speech sample, which was input into the EEG device to synchronize the EEG and EOG signals with the speech stimuli. 

\subsection{EEG analysis.}
The EEG data were processed using MATLAB (MathWorks Inc., USA) and the EEGLAB toolbox \citep{delorme_2004}. After applying a bandpass finite impulse response (FIR) filter between 0.5 Hz and 50 Hz (3300th order), the signals were resampled at 200 Hz. Artifact subspace reconstruction was performed to remove transient, large-amplitude artifacts from the EEG data \citep{mullen_2015}. Further, independent component analysis was applied to the signals, and artifactual components caused by eye movements and blinks were removed from them. Finally, a bandpass FIR filter of 1–8 Hz (1320th order) was used to improve the signal-to-noise ratio. For each speech sample, we prepared a stimulus matrix (i.e., features [moral words, ideological words, policy words, and non-target content words] × time points) at the same sampling rate as the preprocessed EEG data. This matrix comprised time-aligned impulses with a value of 1 at the onset time points of the words and 0 at the other time points. The mTRF toolbox in MATLAB \citep{crosse_2016} was used to estimate the TRF for the content words that described the linear mapping between the stimulus and the preprocessed EEG data. 

The TRF weights over the range of time lags from -100 ms to 1000 ms, relative to the onset of each feature, were estimated using ridge regression with a regularization parameter (lambda = $2^4$). The regularization parameter was determined as follows \citep{ihara_2021}. The mTRFcrossval function \citep{crosse_2016} with leave-one-out cross-validation (LOOCV), in which one trial was evaluated as test data and the remaining trials as training data, was used. The evaluation was repeated until all trials were used as test data. For each participant’s channel data, TRF weights were estimated for every ridge parameter ($2$, $2^2$, $2^3$, $2^4$, $2^5$, $2^6$, $2^7$, $2^8$, and $2^9$). For each ridge parameter, the single-trial weights were averaged over the training data for each parameter. The mean squared errors (MSEs) between the actual responses and the responses predicted using the averaged weights were computed using the test data. The MSEs were averaged over trials, channels, and participants. Finally, we adopted lambda, such that it gives the lowest MSE as the regularization parameter for all participants.

To identify the component corresponding to N400, the TRFs for all word categories were averaged across participants for each channel, and a negative negative-going deflection was found with a peak at 400 ms, most pronounced at the Pz electrode. Hence, for each participant, we calculated the average amplitude from 300 ms to 700 ms of the TRF weights at the Pz for each word category (i.e., moral, ideological, and policy).

\subsection{Statistical analysis.}

A two-way mixed-design ANOVA with Word (within-subject factor: moral, ideological, and policy) and Group (between-subject factor: left, neutral, and right) was employed on the N400 amplitudes to analyze whether political orientation influenced semantic processing for political words. Mauchly’s test confirmed that the homogeneity of variance was not violated. When a significant interaction of Word $\times$ Group was obtained, a one-way ANOVA with a factor of Group was performed for each word category. When the main effect was statistically significant, multiple comparisons were conducted, in which {\it p}-values were adjusted using the Holm procedure. In addition, as a planned comparison, a paired-t test was conducted for each of the left and right groups to ascertain whether there was a significant difference between the N400 amplitude for moral words and that for policy words. In all statistical test procedures, the significance level was set at 0.05. 

To predict political orientation using the EEG responses, the following regression model was run using ordinary least squares:
\begin{equation}\label{eq:regression}
 \mbox{Political}_i = \alpha + \sum_{k \in \{\mbox{moral}, \mbox{ideological}, \mbox{policy} \}} \beta_k \mbox{EEG responses}_{ki} + \epsilon_i, 
\end{equation}
where {\it $\mbox{Political}_i$} is the self-reported political orientation (7-scale), {\it $\mbox{EEG responses}_{ki}$} indicates the individual $i$'s EEG responses for moral, ideological, and policy words, and $\epsilon_i$ is the error term robust to heteroskedasticity \citep{mackinnon_1985}. The Supplementary Material includes the regression results ({\it SI Appendix}, Table \ref{tab:reg_right_left}). Then, the predicted values were computed using the estimated parameters $\beta_k$'s. Finally, the predicted values were correlated with the self-reported values, and Pearson’s correlation coefficient and the {\it p}-value were reported. The statistical analysis was performed with IBM SPSS Statistics 26.0 (IBM Corp.) and R 4.2.2. \\

\noindent {\bf Acknowledgments.} The study was supported by a Nomura Foundation Grant N18-3-E30-014. We thank Kotaro Fujisaki, Sayaka Kaito, Daijiro Kawanaka, Haruhi Oka, Yuta Shimodaira, Charles Genki Shimokura, and Manami Tsuruta for research assistance.

\bibliographystyle{unsrtnat}
\bibliography{main}

\clearpage

\section*{Figures}

\clearpage

\clearpage

\clearpage

\clearpage

\appendix

\renewcommand{\thesection}{A.\arabic{section}} \setcounter{section}{0} %
\renewcommand{\thefigure}{A.\arabic{figure}} \setcounter{figure}{0} %
\renewcommand{\thetable}{A.\arabic{table}} \setcounter{table}{0}

\title{Supporting Information for {\it Semantic Processing of Political Words in Naturalistic Information Differs by Political Orientation}}\author{Shuhei Kitamura\thanks{%
CiDER, Osaka University. Email: kitamura@cider.osaka-u.ac.jp.} \and Aya S. Ihara}
\date{}

\maketitle

\section{Supporting Text}

\subsection{Selection of Participants.}\label{sec:selection}
\begin{figure}[H]
    \begin{center}
    \includegraphics[width=0.8\linewidth]{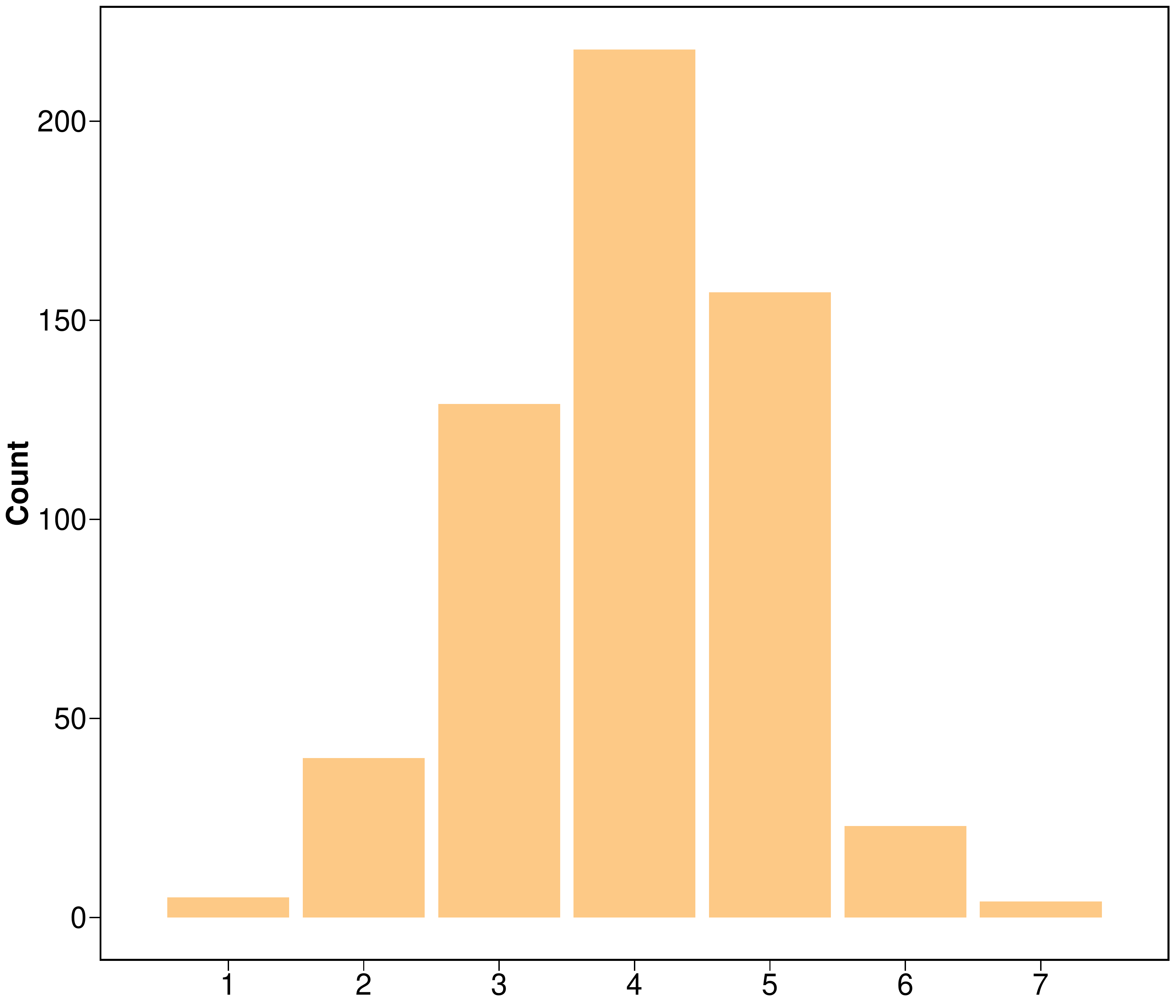}
    \end{center}
    \caption{Distribution of political orientations. The distribution of the political orientation of the 576 individuals who responded to the screening survey.\label{fig:hist}}
\end{figure}

\begin{figure}
    \begin{center}
    \includegraphics[width=0.6\linewidth]{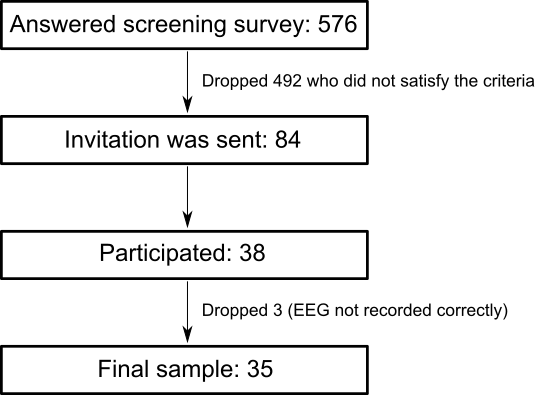}
    \end{center}
    \caption{Sample selection procedure. The numbers in the boxes indicate the number of individuals in each step.\label{fig:selection}}
\end{figure}

A screening survey was sent to all 3,169 students of Osaka University who were registered in the subject pool of the Institute of Social and Economic Research (ISER), Osaka University. In total, 576 people responded to the survey. Then, an invitation email was sent to 84 people (left 30; neutral 29; right 25) who (a) answered either 1 or 2 in ``How interested are you in politics and political issues?" (1 very interested, 2 fairly interested, 3 not very interested, 4 not at all interested) and (b) answered 1 or 2 (left), 4 (neutral), or 6 or 7 (right) in ``People talk of ``the left" and ``the right" in political positions. How would you place your views on this scale, generally speaking?" (Likert scale from left (1) to right (7)) in the survey. Since politically neutral people were the largest group in the sample (Figure \ref{fig:hist}), a subset of them (30 individuals) was randomly selected so that the sample size became more comparable with that of the right and left orientated people. Figure \ref{fig:selection} summarizes the sample selection procedure.

\subsection{Understanding of the Content of Audio Stimuli.}
\begin{figure}
    \begin{center}
    \includegraphics[width=0.8\textwidth]{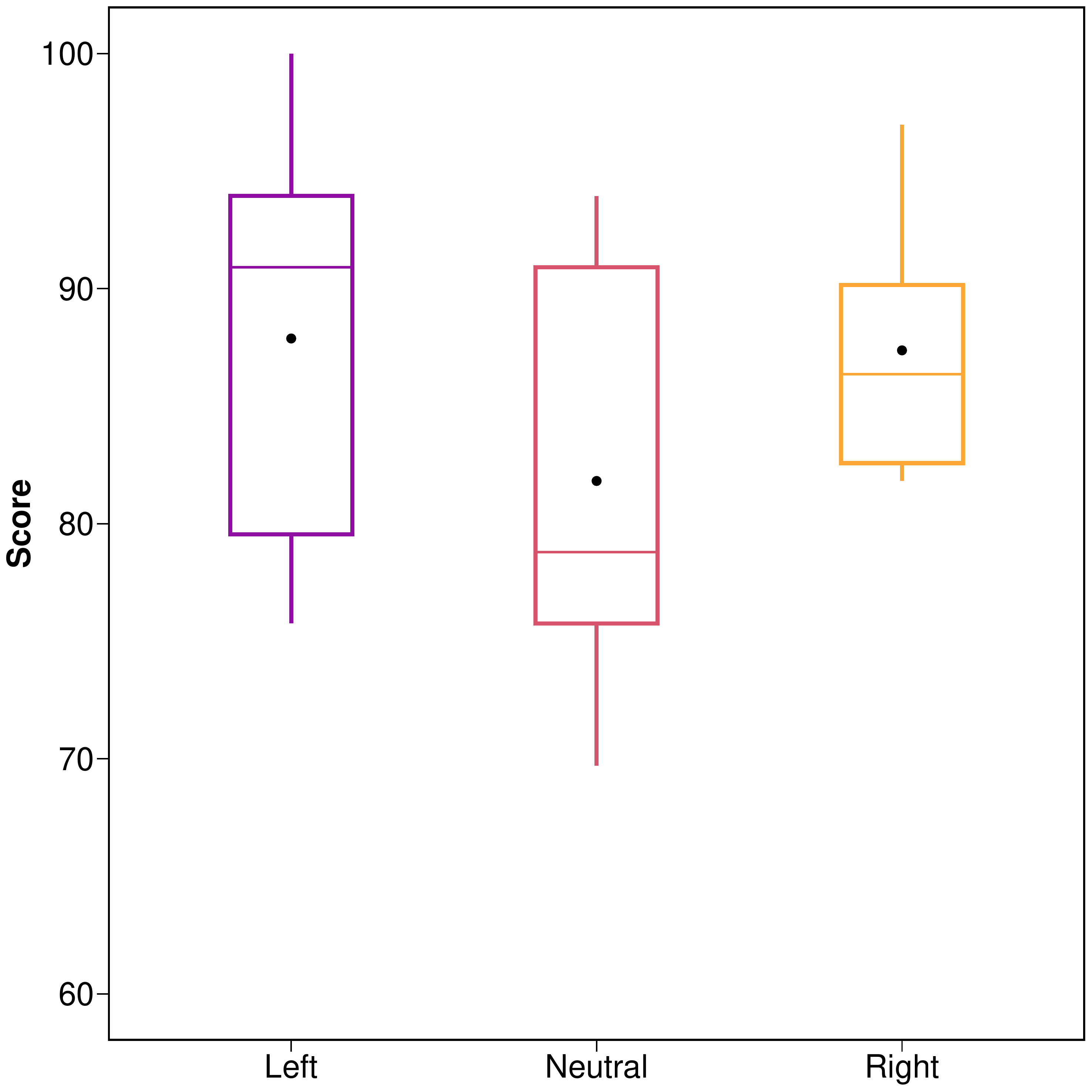}
    \caption{Percentage of correct answers in the choice task. The percentage of correct answers was computed for each individual. Then the distribution of these values for the political right (Right), neutral (Neutral), and left (Left) groups was plotted. \label{fig:correct}}
    \end{center}
\end{figure}

The participants did a choice task after three to seven sentences so that they would pay attention to the speech stimuli. The percentage of correct answers in the task was computed for each individual to investigate the understanding of the content of the audio stimuli. Wilcoxon rank sum tests, with corrections for multiple testing (the Holm procedure), were conducted. We found that individuals with different orientations correctly understood the content of the audio stimuli equally well (left vs. neutral: $p$ = 0.615; left vs. right: $p$ = 0.934; neutral vs. right: $p$ = 0.718) (Figure \ref{fig:correct}).

\subsection{Familiarity of Words.}
To assess the difference in the familiarity of words in the three word categories, we conducted a follow-up survey in November 2021. Since some participants were already dropped from the subject pool of the ISER for reasons such as graduation, we were able to reach only 20 individuals, 15 of whom responded to the survey. One individual was dropped because the person did not complete the survey. Thus, the final sample was 14 individuals.

\begin{figure}
    \begin{center}
    \includegraphics[width=0.8\textwidth]{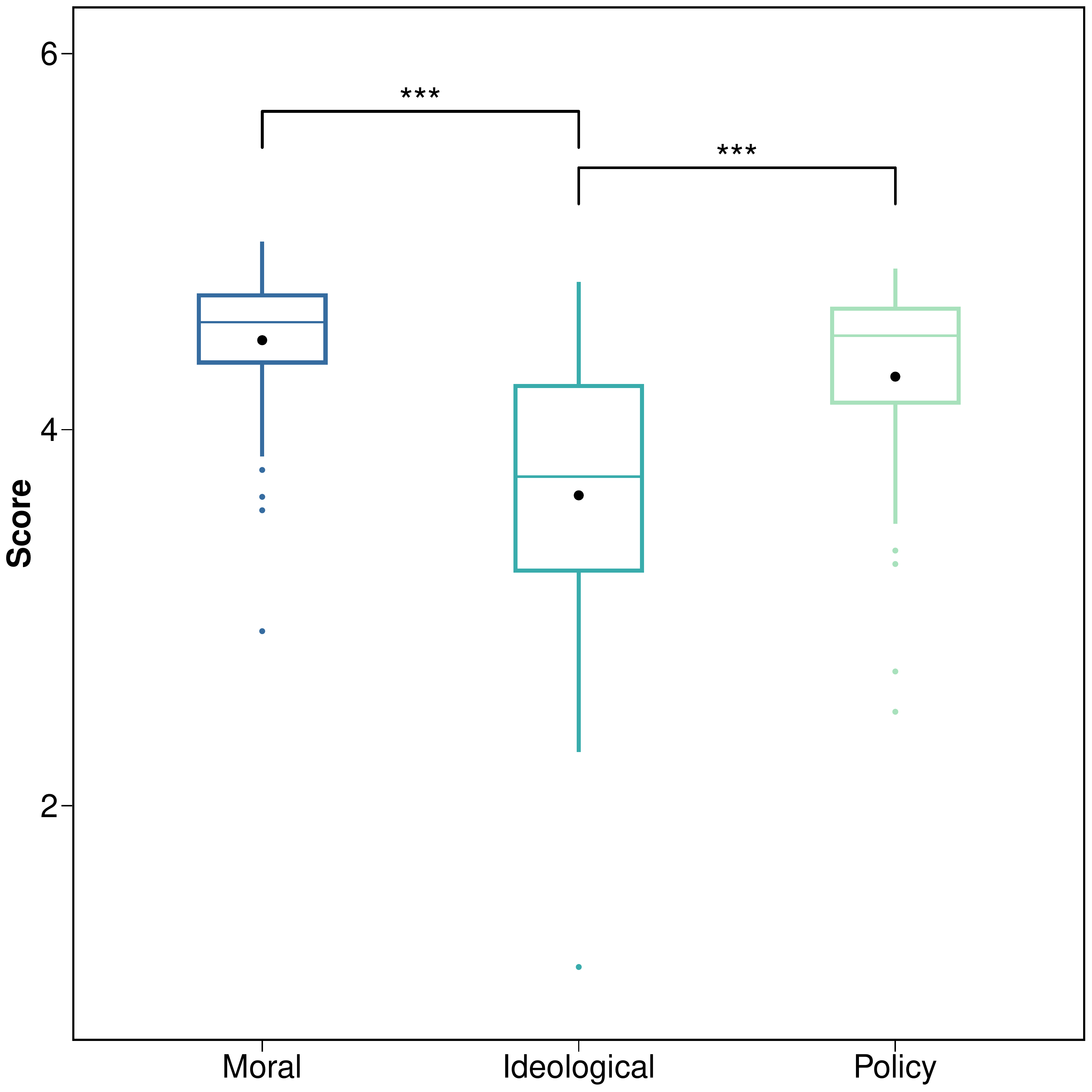}
    \caption{Familiarity of words among participants. The average value for each word was calculated by averaging survey answers. Original scores ranged from 1 (``Not familiar at all") to 5 (``Very familiar"). ***$p$ $<$ 0.001.\label{fig:familiarity}}
    \end{center}
\end{figure}

In the survey, we asked them to evaluate the familiarity of each word. The scale ranged from 1 (``Not familiar at all") to 5 (``Very familiar"). Wilcoxon rank sum tests, with corrections for multiple testing (the Holm procedure), were performed. We found that participants were less familiar with ideological words than moral words ($p$ = 0.000) and policy words ($p$ = 0.000), but the familiarity with moral words and policy words was similar ($p$ = 0.073) (Figure \ref{fig:familiarity}).

\clearpage

\section{Supporting Figures}

\vspace{6cm}

\clearpage

\clearpage

\clearpage

\begin{figure}
    \begin{center}
    \includegraphics[width=\textwidth]{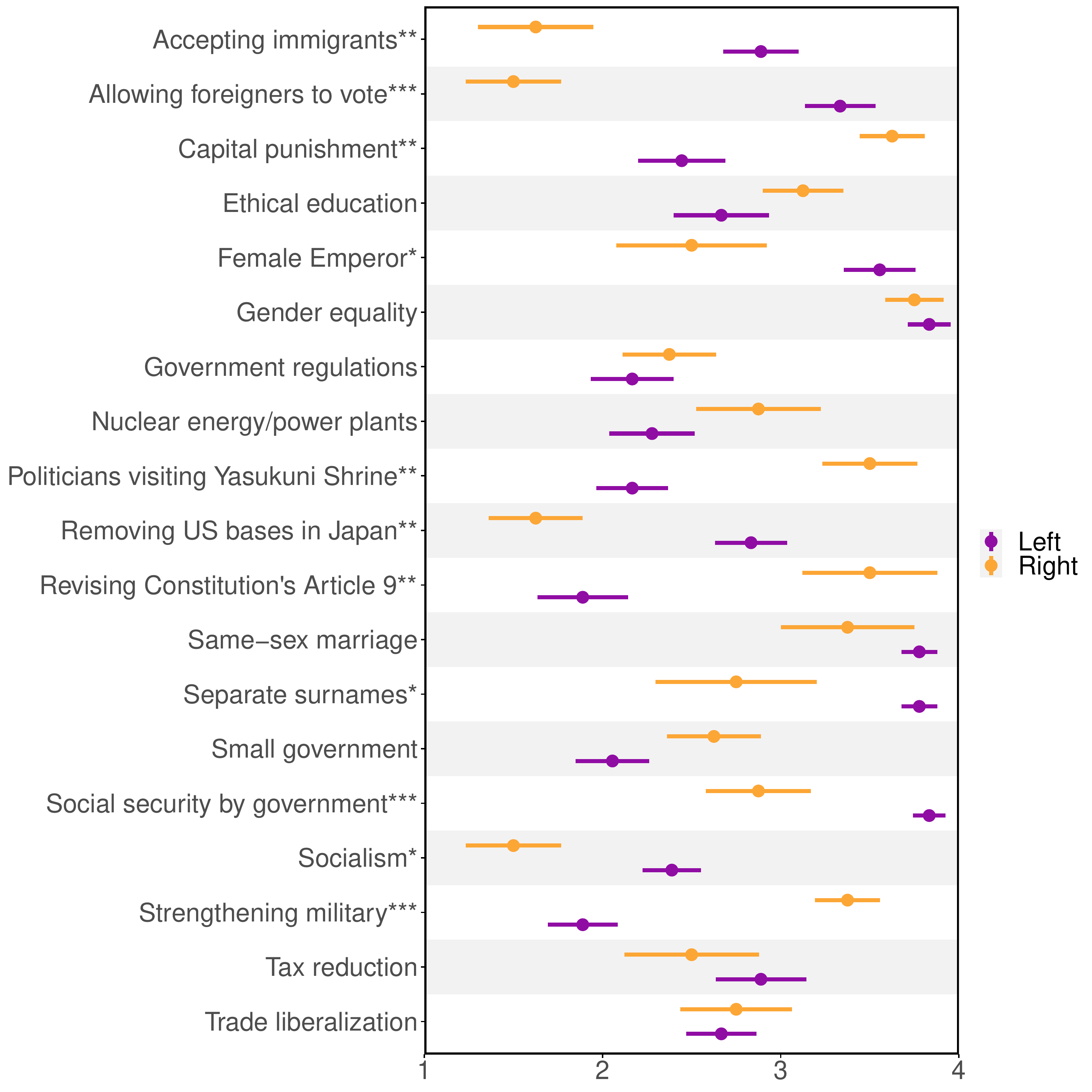}
    \caption{Attitudes toward policies between participants with different political orientations. The scale ranged from 1 (``I disagree") to 4 (``I agree"). *$p$ $<$ 0.05, **$<$ 0.01, and ***$<$ 0.001.\label{fig:poli_attitudes}}
    \end{center}
\end{figure}


\clearpage

\section{Supporting Tables}

\vspace{6cm}

\begin{table}[h]
\begin{center}
\footnotesize
\begin{threeparttable}
\begin{tabularx}{17cm}{@{}X*{9}{D{.}{.}{4.4}}} 
\toprule
&\multicolumn{1}{c}{Left} &\multicolumn{1}{c}{Neutral}   &\multicolumn{1}{c}{Right}      &\multicolumn{2}{c}{Left vs. Neutral} &\multicolumn{2}{c}{Left vs. Right} &\multicolumn{2}{c}{Neutral vs. Right} \\ \cmidrule(lr){5-6} \cmidrule(lr){7-8} \cmidrule(lr){9-10}
&&&&\multicolumn{1}{c}{t-test}&\multicolumn{1}{c}{Wilcox} &\multicolumn{1}{c}{t-test}&\multicolumn{1}{c}{Wilcox}&\multicolumn{1}{c}{t-test}&\multicolumn{1}{c}{Wilcox} \\ 
\midrule 
Age & 25.22 & 22.25 & 21.89 & 0.71 & 0.63 & 0.71 & 0.31 & 0.91 & 0.30  \\
   & (9.14) & (1.58) & (1.90) \\ 
Male & 0.61 & 0.5 & 0.89 & 0.58 & 0.87 & 0.32 & 0.73 & 0.30 & 0.87 \\
   & (0.50) & (0.54) & (0.33) \\ \midrule \addlinespace[0.2cm]
N  & 18 & 8 & 9  \\
\bottomrule
\end{tabularx}
 \end{threeparttable}
\end{center}
\caption{Balance checks. ``t-test" indicates the {\it p}-value using the t-test with corrections for multiple testing (the Holm procedure). ``Wilcox" means the {\it p}-value using the Wilcoxon rank sum test with corrections for multiple testing (the Holm procedure). \label{tab:balance}}
\end{table}

\clearpage

\begin{table}
\begin{center}
\footnotesize
\begin{threeparttable}
\begin{tabularx}{11cm}{@{}X*{4}{D{.}{.}{4.4}}} 
\toprule 
  & \multicolumn{4}{c}{Dependent variable: Political orientation} \\ \cmidrule(lr){2-5}
                     &\multicolumn{1}{c}{(1) } &\multicolumn{1}{c}{(2)}&\multicolumn{1}{c}{(3)} &\multicolumn{1}{c}{(4)} \\ \midrule
Moral N400 Pz       &      -4.251      &                  &                  &      -4.089      \\
                    &     (1.288)^{**} &                  &                  &     (1.400)^{**} \\
Ideological N400 Pz    &                  &       1.351      &                  &       0.698      \\
                    &                  &     (1.917)      &                  &     (1.751)      \\
Policy N400 Pz       &                  &                  &       1.590      &       0.890      \\
                    &                  &                  &     (1.944)      &     (1.759)      \\ 
constant               &       3.569      &       3.761      &       3.759      &       3.654      \\
                    &     (0.209)^{***}&     (0.257)^{***}&     (0.277)^{***}&     (0.233)^{***}\\ \addlinespace[0.1cm] \midrule 
Dep. var. mean (std)&\multicolumn{1}{c}{3.69}      &\multicolumn{1}{c}{3.69}      &\multicolumn{1}{c}{3.69}      &\multicolumn{1}{c}{3.69}      \\
                    &\multicolumn{1}{c}{(1.43)}      &\multicolumn{1}{c}{(1.43)}      &\multicolumn{1}{c}{(1.43)}      &\multicolumn{1}{c}{(1.43)}      \\
R$^2$               &\multicolumn{1}{c}{0.17}      &\multicolumn{1}{c}{0.02}      &\multicolumn{1}{c}{0.01}      &\multicolumn{1}{c}{0.18}      \\
N                   &\multicolumn{1}{c}{35}      &\multicolumn{1}{c}{35}      &\multicolumn{1}{c}{35}      &\multicolumn{1}{c}{35}      \\
 \bottomrule
\end{tabularx}
 \end{threeparttable}
\end{center}
\caption{Ordinary least square regressions of political orientation on EEG responses. Robust standard errors are in parentheses (MacKinnon and White, 1985). The dependent variable is the self-reported political orientation (7-scale). **$p$ $<$ 0.01 and ***$<$ 0.001.\label{tab:reg_right_left}}
\end{table}

\end{document}